\documentclass[12pt]{article}  

\RequirePackage{graphicx,color}
\usepackage{amssymb,amsmath,slashed,mathrsfs,cite}

\usepackage{bbold,parskip}
\newcommand{\gdualn}[1]{\overset{\:{}^{{}^{\boldsymbol{\neg}}}}{\smash[t]{#1}}} 
\newcommand{\0}{\mbox{\boldmath$\displaystyle\mathbb{O}$}}

\newcommand{\J}{\mbox{\boldmath$\displaystyle\boldsymbol{J}$}}

\def\I{\openone}
\def\s{\mbox{\boldmath$\displaystyle\boldsymbol{\sigma}$}}
\def\p{\mbox{\boldmath$\displaystyle\boldsymbol{p}$}}
\def\0{\mbox{\boldmath$\displaystyle\boldsymbol{0}$}}

\def\openone{\mathbb I}

\begin{document}

\date{\empty}

\noindent
\textbf{\large Mass dimension one fields with Wigner degeneracy: A theory of dark matter}

\vspace{11pt}

\noindent
{\textbf{Dharam Vir Ahluwalia}}$^{a,1}$,
\textbf{Julio M. Hoff da Silva}$^{b,2}$,\\
\textbf{Cheng-Yang Lee}$^{c,3}$
\vspace{11pt}
\textcolor{red}{\hrule}
\vspace{11pt}
\begin{quote}
\small{
\noindent $^a$ Center for the Studies of the Glass Bead Game, Notting Hill,\\  Victoria 3168, Australia   \\ 
          $^b$ Departamento de Fisica,  Univeridade Estadual Paulista, UNESP, Guaratinguet\'a, SP, Brazil \\ 
        $^c$   Center for Theoretical Physics, College of Physics, Sichuan University, Chengdu, 610064, China } \\ \\
$^1$ Email: dharam.v.ahluwalia@gmail.com [Corresponding author]\\
 $^2$ E-mail: julio.hoff@unesp.br\\
$^3$ E-mail:  cylee@scu.edu.cn 
\end{quote}

\vspace{11pt}
\begin{quote}
\textbf{Abstract.}
Whatever dark matter is, it must be one  irreducible unitary representation of the extended Lorentz group or another.
We here develop a formalism of mass dimension one fermions and bosons of spin one half, and show that they provide natural dark matter candidates. 
By construction, they are covariant under space-time translations and boosts. However, incorporating the rotational symmetry is non-trivial and requires introducing a two-fold Wigner degeneracy thus doubling the degrees of freedom for particles and anti particles from two to four. With Wigner degeneracy, we have a well-defined theory of mass dimension one fields of spin one half that are physically distinct from the Dirac field. They are local, Lorentz covariant and have positive definite free Hamiltonians. The developed framework also has the potential to resolve the cosmological constant problem, and supply dark energy. 
\vspace{11pt}

\noindent
\textbf{Keywords. } {Elko, Mass dimension one fermions, spin-half bosons, Wigner degeneracy, dark matter, dark energy}

\end{quote}\newpage
\vspace{11pt}
\noindent
\textbf{1. Introduction} \vspace{11pt}

Without exaggeration, dark matter and dark energy reflect an incompleteness of the standard model (SM) of high energy physics. Here, we weave a story in which both of these aspects find a natural solution. We propose that dark matter particles arise as  irreducible unitary representations of the extended Lorentz group with a two-fold Wigner degeneracy~\cite{Wigner:1962ep}.\footnote{Also see Appendix C  of Chapter 2 in~\cite{Weinberg:1995mt}, and the referee reports collected together in the Acknowledgement section of~\cite{Ahluwalia:1995ur}.} 
 Besides the two fold degeneracy, the new fields  are endowed with mass dimension one.
 
  Combined, we find that these facts severely restrict the constructed fields to interact with the SM fields. Thus these  provide  \emph{natural} dark matter candidates. 
Furthermore, the formalism is endowed with not only spin one half fermions but also with spin one half bosons in a unified manner. 
Coupled with recent works~\cite{Ahluwalia:2022dva}, this circumstance cancels the fermionic and bosonic contributions to the vacuum energy density for  the matter fields of the dark as well as SM sector. Equivalence principle then suggests that the quantum corrections to these energy densities must be of the same sign, and thus have the potential to account for the observed dark energy.

The breakthrough presented here realises that attempts to bring the formalism of mass dimension one fields of spin one half in line with that of Weinberg~\cite{Weinberg:1995mt} requires the introduction of a two-fold Wigner degeneracy. When this is done, a formalism at par with that of Weinberg emerges with the consequences noted above.
As observed in the Abstract, the mass dimension one fields are, by construction, covariant under space-time translations and boost. For this reason, we will mainly focus on the constraint introduced by the rotational symmetry, and find a solution that had evaded the subject so far.

\vspace{11pt}
\noindent
\textbf{3. A new quantum field with a two-fold degeneracy}

\vspace{11pt}
All works on mass dimension one fermions are founded on 
the observation that~\cite{ahluwalia_2019,AHLUWALIA20221}:

\begin{quote}
if $\phi(\p)$ transforms as a left-handed Weyl spinor, then up to a multiplicative phase factor $\zeta_\lambda$, the $\Theta \phi^\ast(\p)$ transforms as a right-handed Weyl spinor; and vice versa, that is: 
if $\phi(\p) $ transforms as a right-handed Weyl spinor, then up to a multiplicative phase factor $\zeta_\rho$, the $\Theta \phi^\ast(\p)$ transforms as a left-handed Weyl spinor.
\end{quote}
Here, $\Theta$ is the Wigner time reversal operator for spin one half
\begin{equation}
\Theta = \left(
\begin{array}{cc}
0 & -1\\
1 & 0
\end{array}\right).
\end{equation}
By choosing the spin projection of the left-handed Weyl spinor to be $\pm1/2$ and taking $\zeta_\lambda = + i$ we obtain self conjugate Elko. Setting $\zeta_\lambda = - i$ we get anti self conjugate Elko. Here, Elko is  the German acronym for \textbf{E}igen\-spinoren des \textbf{L}adungs\textbf{k}onjugations\textbf{o}perators. In English, it means eigenspinors of the charge-conjugation operator. The indicated self and anti self conjugacy are under the $\mathcal{R}\oplus\mathcal{L}$ charge conjugation operator
\begin{equation}
\mathcal{C} = 
\left(
\begin{array}{cc}
\0 & i \Theta\\
-i\Theta &\0
\end{array}
\right) K.
\end{equation}
where $K$ complex conjugates functions and spinors to its right. Specifically, self conjugate and anti self conjugate Elko have eigenvalue $1$ and $-1$ with respect to $\mathcal{C}$.

This procedure provides the formalism with four self conjugate and with four  anti self conjugate spinors.
Those constr\-ucted from the left handed Weyl spinors are denoted~\cite{AHLUWALIA20221} by $\lambda^S_+(\p), \lambda^S_-(\p),\lambda^A_+(\p),\lambda^A_-(\p)$ and those constructed from the right handed Weyl spinors are denoted by 
$\rho^S_+(\p), \rho^S_-(\p),\rho^A_+(\p),\rho^A_-(\p)$.

In implementing the rotational invariance of the fields, with Elko as expansion coefficients, the spinors at rest play a fundamental role.  Therefore, we give explicit expressions for the $\lambda^{S,A}_\pm(\0)$
\begin{align}
\lambda^S_+(\0)  &= \sqrt{m}\left(
\begin{array}{cc}
0 \\
 i \\ 
 1 \\
 0
\end{array}
\right),\quad
\lambda^S_-(\0) = \sqrt{m}\left(
\begin{array}{cc}
-i \\
 0 \\ 
 0 \\
 1
\end{array}
\right),\label{eq:restlambda}\\
\lambda^A_+(\0) &= \sqrt{m}\left(
\begin{array}{cc}
i\\
 0 \\ 
 0 \\
 1
\end{array}
\right),\quad
\lambda^A_-(\0) = \sqrt{m}\left(
\begin{array}{cc}
0 \\
 i \\ 
 -1 \\
 0
\end{array}
\right),
\end{align}
and  for the $\rho^{S,A}_\pm(\0)$ 
\begin{align}
\rho^S_+(\0) & = \sqrt{m}\left(
\begin{array}{cc}
1 \\
 0 \\ 
 0 \\
 - i
\end{array}
\right),\quad
\rho^S_-(\0) = \sqrt{m}\left(
\begin{array}{cc}
0 \\
 1 \\ 
 i \\
 0
\end{array}
\right),\\
\rho^A_+(\0) &= \sqrt{m}\left(
\begin{array}{cc}
0\\
 1\\ 
 -i \\
 0
\end{array}
\right),\quad
\rho^A_-(\0) = \sqrt{m}\left(
\begin{array}{cc}
-1\\
 0 \\ 
 0 \\
 -i
\end{array}
\right).\label{eq:restrho}
\end{align}

To introduce the two fold degeneracy, we note that the non linear nature of $\mathcal{C}$ operator transmutes the self and anti self conjugacy under $\lambda(\0) \to \pm i \lambda(\0)$ (and also under 
 $\rho(\0) \to \pm i \rho(\0)$). Thus, besides  $\lambda^S_\pm(\p)$ as self conjugate Elko, we have $-i \lambda^A_\pm(\p) = \rho_\pm^S(\0)$ as self conjugate Elko. Similarly, besides  $\lambda^A_\pm(\p)$ as anti self conjugate Elko, we have $-i \lambda^S_\pm(\p) = \rho^A(\0)$ as anti self conjugate Elko.   This is entirely due to the non-linear nature of $\mathcal{C}$.
 
  It then allows a doubling of the degrees of freedom long ago proposed by Wigner on general quantum field theoretic grounds~\cite{Wigner:1962ep}. Exploiting this circumstance, we define a  self conjugate set of Elko
 \begin{align}
 &\xi(\0,1) = \lambda^S_+(\0),\quad
 \xi(\0,2) = \lambda^S_-(\0),\\
 &\xi(\0,3)  = \rho^S_+(\0),\quad
 \xi(\0,4) = \rho^S_-(\0),
 \end{align}
 and an anti self conjugate set of Elko
\begin{align}
 &\zeta(\0,1) = \lambda^A_+(\0),\quad
  \zeta(\0,2) = \lambda^A_-(\0),\\
& \zeta(\0,3) = \rho^A_+(\0),\quad
 \zeta(\0,4) = \rho^A_- (\0),
 \end{align}
 to define a new quantum field
 \begin{align}
 \lambda(x)  = &\int \frac{d^3p}{(2\pi)^3}
 \frac{1}{\sqrt{2 m E(\p)}}  \nonumber\\
 &  \times \sum_\sigma \Big[ a(\p,\sigma) \xi(\p,\sigma) e^{-i p \cdot x} 
 + b^\dagger(\p,\sigma) \zeta(\p,\sigma) e^{i p \cdot x} \Big] . \label{eq:field}
 \end{align}
 The annihilation and creation operators are left free to satisfy fermionic, or bosonic, statistics. We will attend to this aspect as we proceed.
In the definition of $\lambda(x)$,  $\xi(\p,\sigma)$ is related to the  $\xi(\0,\sigma)$ by the $\mathcal{R}\oplus\mathcal{L}$ boost operator
\begin{align}
\xi(\p,\sigma) = & D(L(p)) \xi(\0,\sigma)\nonumber\\
= &\sqrt{\frac{E+m}{2m}}
\left[
\begin{array}{cc}
\I +\frac{
{\boldsymbol{\sigma}}\cdot{\boldsymbol{p}}}
{E+m} & \mathbb{0} \\
\mathbb{0} & \I -\frac{{\boldsymbol{\sigma}}\cdot{\boldsymbol{p}}}{E+m} 
\end{array}
\right] \xi(\0,\sigma) ,\label{eq:boost1}
\end{align}
with $L(p)$ representing pure boost. Similarly
\begin{equation}
\zeta(\p,\sigma) = D(L(p)) \zeta(\0,\sigma). \label{eq:boost2}
\end{equation}

Introduction of doubly degenerate set of Elko as just done appears for the first time in literature. It has profound consequences for the covariance under inhomogeneous Lorentz group. Since the $\lambda(x)$ respects boost and spacetime translation symmetry as written down, it must be shown that the Elko at rest must emerge from rotational symmetry and must not be assumed in ad hoc manner. But since we have already introduced the Elko at rest we must show that these are consistent with the rotational symmetry.

Towards this end, we note that 
the rotation generators for Elko, 
are written in a basis in  which $\mathcal{J}_z$ is diagonal. 
Thus the explicit form of $\boldsymbol{\mathcal{J}}$ satisfying  the
$\mathfrak{su}(2)$ algebra is
 \begin{align}
 & \mathcal{J}_x =  \frac{1}{2} \left(
 \begin{array}{cc}
 \sigma_x & \mathbb{o} \\
 \mathbb{o} & \sigma_x
 \end{array}
 \right),\quad \mathcal{J}_y =  \frac{1}{2} \left(
 \begin{array}{cc}
 \sigma_y  & \mathbb{o} \\
 \mathbb{o} & \sigma_y
 \end{array}
 \right),\nonumber  \\ 
 &\mathcal{J}_z =  \frac{1}{2} \left(
 \begin{array}{cc}
 \sigma_z & \mathbb{o} \\
 \mathbb{o} & \sigma_z
 \end{array}
 \right).\label{eq:CurlyJ14}
  \end{align}
The irreducibility of the representation is a bit subtle. Here it suffices to mention that discrete symmetries and mass dimensionality of the fields
account for that.
 
The crucial observation now is the following: If the chosen set of Elko that enter the definition of $\lambda(x)$ are to be consistent with rotational symmetry, there must exist an $\mathfrak{su}(2) $ satisfying  set of 
$\J \stackrel{\mathrm {def}}{=} \left(J_x,J_y,J_z\right)$ for spin half,  such that\footnote{See, equations (5.1.25) and (5.1.26) in~\cite{Weinberg:1995mt}.}
  \begin{equation}
 \sum_{{\sigma}^\prime} \xi_{{\ell^\prime}}(\0,\sigma^\prime)\, \J_{{\sigma}^\prime\sigma} = \sum_\ell  \boldsymbol{\mathcal{J}}_{{\ell}^\prime\ell} \,\xi_\ell(\0,\sigma),\label{eq:rotp}
 \end{equation}
 and 
  \begin{equation}
 \sum_{{\sigma^\prime}} \zeta_{{\ell^\prime}}(\0,\sigma^\prime) \,\J^\ast_{{\sigma^\prime}\sigma} = - \sum_\ell \boldsymbol{\mathcal{J}}_{{\ell^\prime}\ell}\, \zeta_\ell(\0,\sigma).\label{eq:rota}
 \end{equation}
 
 If we use the $u(\0,\sigma)$ and $v(\0,\sigma)$ of the Dirac formalism with appropriate phases on the right hand side of above equations, the $\J= \frac{1}{2}\, \s $ solves the rotational constraints.  However, for Elko one necessarily needs to introduce a two fold degeneracy to solve these constraints. 
A somewhat lengthy  calculation reveals that indeed the $\J$ exists, and is given by
\begin{align}
 & J_x= \frac{i}{2}
 \left(
 \begin{array}{cc}
 \mathbb{o} & -\sigma_z \\
 \sigma_z & \mathbb{o}
 \end{array}
 \right),\quad
 J_y= \frac{1}{2}
 \left(
 \begin{array}{cc}
  \sigma_y  & \mathbb{o} \\
  \mathbb{o} &\sigma_y
 \end{array}
 \right),\nonumber\\
 &J_z= \frac{i}{2}
 \left(
 \begin{array}{cc}
 \mathbb{o} & \sigma_x \\
-  \sigma_x & \mathbb{o}
 \end{array}
 \right).\label{eq:RomanJ17}
 \end{align}
 Furthermore, these indeed satisfy $ \left[J_x,J_y\right]= i J_z$ and cyclic permutations; that is: the $\mathfrak{su}(2)$ algebra. 
We thus have a fully Lorentz covariant quantum field with Elko as its expansion coefficients. The spacetime translation symmetry is implemented through the exponentials
 $e^{\pm i p\cdot  x}$ that appear in $\lambda(x) $, the boost symmetry enters the formalism through (\ref{eq:boost1}) 
 and  (\ref{eq:boost2}) provided $\xi(\0,\sigma)$ and $\zeta(\0,\sigma)$ satisfy the rotational constraint -- something which we have just proven.

We emphasise that without invoking the two fold Wigner degeneracy, the rotational constraints (\ref{eq:rotp}) and (\ref{eq:rota}), with  $\xi(\0,\sigma)$ and $\zeta(\0,\sigma)$ as input, \textit{cannot} be satisfied. So, $\lambda(x)$ is necessarily irreducible. This last assertion is fully supported by an examination of the discrete symmetries, and will be the subject of an archival version of this communication. In the formalism we have developed a choice of  phases for Elko  in 
equations (\ref{eq:restlambda}) to (\ref{eq:restrho}) is incorporated implicitly.\footnote{The phases are restricted to $\pm 1$ to preserve their character under charge conjugation operator $\mathcal{C}$.} Without making this choice, the rotational constraints are violated! Examples of this mistake are wide spread in the Physics and Mathematics  literature  on quantum field theory, see for example the rest spinors for the Dirac formalism presented in  references~\cite{ryder_1996,Folland:2008zz}.

But, both the $\boldsymbol{\mathcal{J}}$ and the $\boldsymbol{J}$, given by equations (\ref{eq:CurlyJ14}) and (\ref{eq:RomanJ17}) respectively, 
 satisfy the same $\mathfrak{su}(2)$ algebra. Consequently, we can apply a similarity transformation $S$ to make $\boldsymbol{J}$ coincide with  $\boldsymbol{\mathcal{J}}$. The $S$ is found not to commute with the discrete symmetries of $\lambda(x)$. 
Thus, there exist at least two physically distinct solutions. One 
an irreducible field 
 $\lambda(x)$, and the other a direct sum of two Dirac fields. The doublet of Dirac fields, not necessarily with the same mass, emerges as a natural companion to $\lambda(x)$. Once the latter is identified with the dark matter sector, the formalism leads to a unification 
of the matter sectors of the SM and the dark sector. 


 \vspace{11pt}
 
 \noindent
\textbf{4. Elko dual, and adjoint}
 \vspace{11pt}

Under the usual Dirac dual, Elko have vanishing norms. For this reason, we introduce the following duals\begin{align}
\gdualn{\xi}(\p,\sigma) & = \left[ + \mathcal{D} \,\xi(\p,\sigma) \right]^\dagger \gamma_0, \\
\gdualn{\zeta}(\p,\sigma) & =  s\left[ - \mathcal{D} \,\zeta(\p,\sigma) \right]^\dagger \gamma_0,
\end{align}
where the Dirac operator $\mathcal{D}$ is defined as
\begin{equation}
\mathcal{D} \stackrel{\mathrm{def}}{=} m^{-1} \gamma_\mu p^\mu.
\end{equation}
Explicit calculations yield the action of $\mathcal{D}$ on Elko, 
and read~\cite{Dvoeglazov:1995eg,Speranca:2013hqa,ahluwalia_2019}
\begin{align}
\mathcal{D}\,\xi(\p,1) = +i \xi(\p,2), \quad  \mathcal{D}\,\xi(\p,2) 
= - i \xi(\p,1),\nonumber\\
\mathcal{D}\,\xi(\p,3)  = - i \xi(\p,4), \quad \mathcal{D}\,\xi(\p,4)=+i \xi(\p,3),
\end{align}
and
\begin{align}
\mathcal{D}\,\zeta(\p,1) &=- i \zeta(\p,2), \quad \mathcal{D}\,\zeta(\p,2) =  + i \zeta(\p,1),\nonumber\\
\mathcal{D}\,\zeta(\p,3) & =  +i \zeta(\p,4),\quad \mathcal{D}\,\zeta(\p,4) = - i \zeta(\p,3).
\end{align}
This property of Elko under $\mathcal{D}$ is crucial to the mass dimensionality of $\lambda(x)$. 
With the duals thus defined, we obtain the orthonormality relations
\begin{equation}
\gdualn{\xi}(\p,\sigma) {\xi}(\p,\sigma^\prime) = 2 m \,\delta_{\sigma{\sigma}^\prime},\;
\gdualn{\zeta}(\p,\sigma) {\zeta}(\p,\sigma^\prime) = - 2 m\,s \,\delta_{\sigma{\sigma}^\prime},\nonumber
\end{equation}
and Lorentz invariant spin sums
 \begin{align}
  \sum_{\sigma} \xi(\p,\sigma) \gdualn{\xi}(\p,\sigma) & = 2m\, \I_4,
 \\
 \sum_{\sigma} \zeta(\p,\sigma) \gdualn{\zeta}(\p,\sigma) & =
 -  2m s \,\I_4 .
 \end{align}
Therefore, the adjoint of $\lambda(x)$ is defined as
 \begin{align}
 \gdualn{\lambda}(x) = &\int \frac{d^3p}{(2\pi)^3}
 \frac{1}{\sqrt{2 m E(\p)}}  \nonumber\\
 &  \times \sum_\sigma \Big[ a^\dagger(\p,\sigma) \gdualn{\xi}(\p,\sigma) e^{i p \cdot x} 
 + b(\p,\sigma) \gdualn{\zeta}(\p,\sigma) e^{-i p \cdot x} \Big] .\nonumber
 \end{align}
\vspace{11pt}

\noindent
\textbf{5. The mass dimension of  $\boldsymbol{\lambda(x)}$ and darkness}
\vspace{11pt}

 Detailed calculations following reference~\cite{AHLUWALIA20221} show that for $s=+ 1$ we obtain a fermionic field, and for $s=-1$ we secure a bosonic field, with annihilation and creation operators satisfying the appropriate anti-commutation or commutation relations. Evaluating the time ordered product\footnote{There is an obvious typo in equation (63) of reference ~\cite{AHLUWALIA20221} . The adjoint symbol is missing  from $\mathfrak{f}$.}
 \begin{equation}
\langle ~\vert\mathfrak{T}  
[ \lambda(x^\prime) \gdualn{\lambda}(x)]
 \vert~\rangle
 \end{equation}
 yields the Feynman-Dyson propagator (for the fermionic as well as the bosonic field, taking into account the statistics parameter $s$ and the above obtained spin sums)
 \begin{equation}
\langle ~\vert\mathfrak{T}  
[ \lambda(x^\prime) \gdualn{\lambda}(x)]
 \vert~\rangle = i\int \frac{d^4p\,}{(2\pi)^{4}}
 e^{-i p\cdot(x^\prime - x)}
\frac{2\I_4}{p_\mu p^\mu - m^2 + i\epsilon} 
 \end{equation}
 Thus mass dimension of the $\lambda(x)$-$\gdualn{\lambda}(x)$ system is one, with the free Lagrangian density given by
 \begin{equation}
 \mathcal{L}_0(x) = \frac{1}{2}\left[
 \partial^\mu\gdualn{\lambda}(x) \partial_\mu\lambda(x)
 - m^2 \gdualn{\lambda}(x) \lambda(x) \right]
 \end{equation}
 With $\mathcal{L}_0(x)$ as above, one can immediately calculate the momentum conjugate to $\lambda(x)$
 \begin{equation}
 \mathfrak{p}(x) = \frac{\partial \mathcal{L}_0(x)}{\partial\dot\lambda(x)} = \frac{1}{2}
 \frac{\partial}{\partial t}
 {\gdualn{\lambda}(x)},
 \end{equation}
 and one can verify the standard locality anti-commutators (for fermions) and commutators (for bosons). Furthermore, modulo vacuum energies, the Hamiltonian is positive definite for fermions as well as bosons; and if they carry the same mass, the contributions to zero point vacuum energies cancel.
 

 


Because of mismatch between mass dimensionalities of one versus three half,
$\lambda(x) $ cannot form doublets with the SM fields. In addition, the presence of
 singlets of the form $\phi(x) \gdualn{\lambda}(x) \psi(x)$, would lead to violation of charge conservation, assuming that dark matter does not carry SM charges.\footnote{$\psi(x)$ is a Dirac field.} This fact renders $\lambda(x)$ a natural dark matter candidate. 
 The allowed interactions are: a quartic self interaction $[\gdualn{\lambda}(x)\lambda(x)]^2$, and  with Higgs via
 $\phi(x)\gdualn{\lambda}(x)\lambda(x)$, and 
 $\phi^2(x)\gdualn{\lambda}(x)\lambda(x)$.

\vspace{11pt}

\noindent
\textbf{6. Discussion and concluding remarks} 
\vspace{11pt}

Mass dimension one fermions  \emph{without} Wigner degen\-eracy now has two decades of history~\cite{Ahluwalia:2004ab,Ahluwalia:2004sz}. Its expansion coefficients, Elko, along with the associated fields, have been widely studied in cosmology to mathematical physics~\cite{Bahamonde:2017ize,Boehmer:2010ma,Lima:2022vrc,Bernardini:2012sc,refId0,Liu:2011nb}.\footnote{For an extensive list of references one may consult~\cite{AHLUWALIA20221}.}
However, from early on, a problem had persisted about formalism's rotational symmetry~
\cite{Gillard:2010nr,Ahluwalia:2008xi,Ahluwalia:2009rh,Ahluwalia:2010zn} till it was resolved by introducing the notion of  $\tau$-deformation in references~\cite{Lee:2014opa,Ahluwalia:2016rwl,Rogerio:2016mxi}.
Here, the problem is manifestly resolved by incorporating Wigner degeneracy. 
 This degeneracy was not pursued by Weinberg on the grounds that~\cite{Weinberg:1995mt}, ``no examples are known of particles that furnish unconventional representations.'' Despite that, one  of the present authors had taken issue with this omission in a book review~\cite{Ahluwalia:1997ga}, and even he did not fully realise the cost of this omission. This paper corrects that now, and 
 incorporates 
  Wigner degeneracy in the formalism -- to the best of our knowledge for the first time.
In a forthcoming paper we plan to report on discrete symmetries and examine in detail the irreducibility of $\lambda(x)$.

We opened this communication with the assertion that 
whatever dark matter is, it must be one  irreducible unitary representation of the extended Lorentz group or another. Here, we have confirmed that if Wigner degeneracy is incorporated in the formalism of mass dimension one fields then indeed dark matter is naturally incorporated in the physics beyond the standard model.

\noindent
\textbf{Acknowledgements}

DVA is supported by the Center for the Studies of the Glass Bead Game, JMHS is supported by CNPq grant No. 303561/2018-1, and CYL is supported by the  Sichuan University Postdoctoral Research Fund No.2022SCU12119




\begin{thebibliography}{10}

\bibitem{Wigner:1962ep}
E.~P. Wigner, {\it {Unitary representations of the inhomogeneous Lorentz group
  including reflection}},  in {\em {Group theoretical concepts and methods in
  elementary particle Physics: Lectures of the Istanbul Summer School of
  theoretical physics, 1962, edited by F. Gursey}}, pp.~564--607, Gordon and
  Breach, 1964.

\bibitem{Weinberg:1995mt}
S.~Weinberg, {\em {The Quantum theory of fields. Vol. 1: Foundations}}.
\newblock Cambridge University Press, 6, 2005.

\bibitem{Ahluwalia:1995ur}
D.~V. Ahluwalia, {\it {A New type of massive spin one boson: And Its relation
  with Maxwell equations}},  {\em Fundam. Theor. Phys.} {\bf 80} (1997)
  443--457, [\href{http://xxx.lanl.gov/abs/hep-th/9509116}{{\tt
  hep-th/9509116}}].

\bibitem{Ahluwalia:2022dva}
D.~V. Ahluwalia and C.-Y. Lee, {\it Spin-half bosons with mass dimension
  three-half: Evading the spin-statistics theorem},  {\em Europhysics Letters
  (EPL)} {\bf 140} (oct, 2022) 24001.

\bibitem{ahluwalia_2019}
D.~V. Ahluwalia, {\em Mass Dimension One Fermions}.
\newblock Cambridge Monographs on Mathematical Physics. Cambridge University
  Press, 2019.

\bibitem{AHLUWALIA20221}
D.~V. Ahluwalia, J.~M. {Hoff da Silva}, C.-Y. Lee, Y.-X. Liu, S.~H. Pereira,
  and M.~M. Sorkhi, {\it Mass dimension one fermions: Constructing darkness},
  {\em Physics Reports} {\bf 967} (2022) 1--43. Mass dimension one fermions:
  Constructing darkness.

\bibitem{ryder_1996}
L.~H. Ryder, {\em Quantum Field Theory}.
\newblock Cambridge University Press, 2~ed., 1996.

\bibitem{Folland:2008zz}
G.~B. Folland, {\em {Quantum field theory: A tourist guide for
  mathematicians}}.
\newblock American Mathematical Society, 2008.

\bibitem{Dvoeglazov:1995eg}
V.~V. Dvoeglazov, {\it {Lagrangian for the Majorana-Ahluwalia construct}},
  {\em Nuovo Cim. A} {\bf 108} (1995) 1467--1476.

\bibitem{Speranca:2013hqa}
L.~D. Speran\c{c}a, {\it {An Identification of the Dirac Operator with the
  Parity Operator}},  {\em Int. J. Mod. Phys. D} {\bf 23} (2014), no.~14
  1444003.

\bibitem{Ahluwalia:2004ab}
D.~V. Ahluwalia and D.~Grumiller, {\it {Spin half fermions with mass dimension
  one: Theory, phenomenology, and dark matter}},  {\em JCAP} {\bf 07} (2005)
  012.

\bibitem{Ahluwalia:2004sz}
D.~V. Ahluwalia and D.~Grumiller, {\it {Dark matter: A Spin one half fermion
  field with mass dimension one?}},  {\em Phys. Rev. D} {\bf 72} (2005) 067701.

\bibitem{Bahamonde:2017ize}
S.~Bahamonde, C.~G. B\"ohmer, S.~Carloni, E.~J. Copeland, W.~Fang, and
  N.~Tamanini, {\it {Dynamical systems applied to cosmology: dark energy and
  modified gravity}},  {\em Phys. Rept.} {\bf 775-777} (2018) 1--122.

\bibitem{Boehmer:2010ma}
C.~G. Boehmer, J.~Burnett, D.~F. Mota, and D.~J. Shaw, {\it {Dark spinor models
  in gravitation and cosmology}},  {\em JHEP} {\bf 07} (2010) 053.

\bibitem{Lima:2022vrc}
R.~d.~C. Lima, T.~M. Guimar\~aes, and S.~H. Pereira, {\it {A pilot study on
  canonical gravity with mass dimension one fermions}},  {\em JHEP} {\bf 09}
  (2022) 132.

\bibitem{Bernardini:2012sc}
A.~E. Bernardini and R.~da~Rocha, {\it {Dynamical dispersion relation for ELKO
  dark spinor fields}},  {\em Phys. Lett. B} {\bf 717} (2012) 238--241.

\bibitem{refId0}
{Moazzen Sorkhi, Masoumeh} and {Ghalenovi, Zahra}, {\it Localization of elko
  spinor fields in tachyonic de sitter braneworld models},  {\em Eur. Phys. J.
  C} {\bf 80} (2020), no.~4 314.

\bibitem{Liu:2011nb}
Y.-X. Liu, X.-N. Zhou, K.~Yang, and F.-W. Chen, {\it {Localization of 5D Elko
  Spinors on Minkowski Branes}},  {\em Phys. Rev. D} {\bf 86} (2012) 064012.

\bibitem{Gillard:2010nr}
A.~Gillard and B.~Martin, {\it {Dark matter, Elko fields and Weinberg's quantum
  field theory formalism}},  {\em Rept. Math. Phys.} {\bf 69} (2012) 113--129.

\bibitem{Ahluwalia:2008xi}
D.~V. Ahluwalia, C.-Y. Lee, and D.~Schritt, {\it {Elko as self-interacting
  fermionic dark matter with axis of locality}},  {\em Phys. Lett. B} {\bf 687}
  (2010) 248--252.

\bibitem{Ahluwalia:2009rh}
D.~V. Ahluwalia, C.-Y. Lee, and D.~Schritt, {\it {Self-interacting Elko dark
  matter with an axis of locality}},  {\em Phys. Rev. D} {\bf 83} (2011)
  065017.

\bibitem{Ahluwalia:2010zn}
D.~V. Ahluwalia and S.~P. Horvath, {\it {Very special relativity as relativity
  of dark matter: The Elko connection}},  {\em JHEP} {\bf 11} (2010) 078.

\bibitem{Lee:2014opa}
C.-Y. Lee, {\it {A Lagrangian for mass dimension one fermionic dark matter}},
  {\em Phys. Lett. B} {\bf 760} (2016) 164--169.

\bibitem{Ahluwalia:2016rwl}
D.~V. Ahluwalia, {\it {The theory of local mass dimension one fermions of spin
  one half}},  {\em Adv. Appl. Clifford Algebras} {\bf 27} (2017), no.~3
  2247--2285.

\bibitem{Rogerio:2016mxi}
R.~J.~B. Rog\'erio and J.~M. Hoff~da Silva, {\it {The local vicinity of spins
  sum for certain mass dimension one spinors}},  {\em EPL} {\bf 118} (2017),
  no.~1 10003.

\bibitem{Ahluwalia:1997ga}
D.~V. Ahluwalia, {\it {Book Review of The Quantum Theory of Fields (by Steven
  Weinberg)}},  {\em Found. Phys.} {\bf 10} (1997) 301--304.

\end{thebibliography}

\providecommand{\href}[2]{#2}\begingroup\raggedright\endgroup

\end{document}